%% file: paper.tex
\RequirePackage{fix-cm}

\documentclass[smallcondensed]{svjour3}     
\smartqed  
\usepackage{graphicx}
\usepackage{url}
\usepackage{xcolor}
\begin{document}

\title{LincoSim: a web based HPC-cloud platform for automatic virtual towing tank analysis
}


\author{R. Ponzini \and
        F. Salvadore 
}


\institute{R. Ponzini \at CINECA, HPC Department
    \email{r.ponzini@cineca.it}           
    \and
    F. Salvadore \at CINECA, HPC Department 
    \email{f.salvadore@cineca.it}
}


\maketitle

\input{abstract}

\input{intro}
\input{overview}
\input{usage}
\input{web_details}
\input{cfd_details}
\input{conclusions}
\input{credits}

\bibliographystyle{spphys}       
\bibliography{mybib}   


\end{document}

%% file: abstract.tex
\begin{abstract}
Thanks to evolving web technologies, computational platforms, automation tools and open-source software business model, today, it is possible to develop powerful automatic and virtualized web services for complex physical problems in engineering and design.

In particular, in this work, we are presenting a new web based HPC-cloud platform for automatic virtual towing tank analysis.
It is well known that the design project of a new hull requires a continuous integration of shape hypothesis and hydrodynamics verifications using analytical tools, 3D computational methods, experimental facilities and sea keeping trial tests. The complexity and the cost of such design tools increase considerably moving from analytical tools to sea keeping trials. In order to perform a meaningful trade-off between costs and high quality data acquiring during the last decade the usage of 3D computational models has grown pushed also by well-known technological factors. 

Nevertheless, in the past, there were several limiting factors on the wide diffusion of 3D computational models to perform virtual towing tank data acquiring. On one hand software licensing and hardware infrastructure costs, on the other hand the need of very specific technological skills limited the usage of such virtualized tools only to research centers and or to large industrial companies. In this work we propose an innovative
    high-level approach which is embodied in the so-called \textit{LincoSim} (\cite{lincosim}) 
    web application in which a hypothetical \textit{designer user} can
    carry out the simulation only starting from its own geometry and a set of meaningful physical parameters. \textit{LincoSim} automatically manages and hides to the user all the necessary details of CFD modelling and of HPC infrastructure usage allowing them to access,
visualize and analyze the outputs from the same single
access point made up from the web browser. In addition to the web interface, the platform includes a back-end server which implements a Cloud logic and can be
connected to multiple HPC machines for computing.
    \textit{LincoSim} is currently set up with finite volume Open-FOAM CFD engine. A preliminary validation campaign has been performed to assess the robustness and the reliability of the tool and  is proposed as a novel approach for the development of Computer Aided
Engineering (CAE) applications.

\keywords{High Performance Computing \and Cloud \and Computational Fluid Dynamics \and Naval Engineering \and OpenFOAM \and Design \and }

\end{abstract}

%% file: intro.tex
\section{Introduction}
\label{intro}

Ship design involves a high number of variables and requirements that require remarkable effort
for managing the complexity of interactions. An integral approach is to date the most 
ambitious objective (for a review see \cite{papanikolaou2010}) and the combination of established 
and innovative methods allows to get better and better results. For several decades, the design 
process has been also computer-aided, always evolving as methodologies and tools (see \cite{nowacki2010}). 
In this context, the integration of application software plays a significant role (see \cite{tampier2014}).

Among multiple challenges, one of the major 
challenges is to enhance the loop of information between product design and preliminary performance evaluation.  
In the typical design process, where real experiments are used, diverse major issues concerning the integration 
loop are evident:
\begin{itemize}
    \item high time to results: easily weeks or months from design concept to Key Parameter Indices (KPI) acquiring;
    \item high costs: mostly related to physical prototype construction rather than to the renting of the experimental facility itself;
    \item poor integration with  a design loop activity: after a minimal design update or working condition change the overall experimental setup must be rebuild from scratch or so;
    \item unfeasible integration with existing internet of things (IoT) facilities and concepts.
\end{itemize}

In this context, the Computational Fluid Dynamics (CFD) can be the building brick for a \textit{virtual experiment}. The objective is to assess relevant hydrodynamics features thus fixing the lack of information of real product usage by providing tools to support continuous products improvement cycles through tailored simulations. This approach can help the vessels producers to reach their business goals in a more efficient, effective and in a shorter time to market way, increasing companies' competitiveness. However, a CFD based approach has also critical issues. Mainly, to get good quality results advanced numerical and software skills are required while, to get the results within reasonable times, specialized hardware may be needed, i.e. High Performance Computing (HPC) machines. In addition to the need for accessing the machines, the usage of HPC systems also requires specific skills which must be considered when planning an integrated process design workflow.

In our vision, a \textit{virtual towing tank} simulation facility can be considered as winning or effective in an integrated process design if it is:
\begin{itemize}
    \item interactively usable by a non-expert user;
    \item automatic and transparent respect to CFD modelling concepts and HPC facility deploy;
    \item robust to different hull shapes and working conditions;
    \item effective respect to required KPI generally used in hull hydrodynamics computations.
\end{itemize}

We designed an innovative application named \textit{LincoSim} to be usable by a so-called \textit{designer user} that represents a domain expert in hull design but not necessarily able to manage a complete CFD workflow or able to use an HPC facility.
More in detail, in this \textit{virtual towing tank} there are two main building blocks: the user interface and the computational engine. The first block -- entirely web-based -- allows the user to interact with the virtual experimental facility in order to provide the inputs to the system, to visualize and to check the formal correctness of the desired working condition to be tested, submit the analysis, visualize, navigate and compare the results provided by the system. The second block is the undergoing computational workflow that is automatically instructed by the users requests and that emulates the physics of hull undergoing a well defined working condition and that is able to autonomously compute and return a set of standardized KPI that are generally available after a real towing tank session.

The \textit{LincoSim} web application is part of the LINCOLN project is a solution that includes both these two blocks using a fully open-source based software stack including the possibility to transparently use HPC facilities.
LincoSim is developed as a part of the LINCOLN project (\cite{lincoln}) which is a complex project where innovative vessels are designed according to lean design tools (KbeML – Knowledge Based Engineering Modelling Language) and methodologies (SBCE – Set Based Concurrent Engineering), taking care of sustainability of the whole process, from environmental (LCA - Life Cycle Assessment) and financial (LCC - Life Cycle Cost) point of view and adopting digital solutions, through an integrated IoT (Internet of Things) platform, able to provide knowledge and future services to the maritime sector actors.

Moving software interfaces towards the web access is a common trend nowadays and it is becoming popular
also in the context of Computer-Aided Engineering applications (see e.g., \cite{onshape} or \cite{simscale}).
The associated cloud design allows to distribute data and computing in a very natural and efficient way.
A more advanced approach is not only to provide web interfaces to existing applications but to give the
chance to access complete work-flows. \textit{LincoSim} somehow inherits ideas from this Workflow as a Service (WfaaS) approach and a 
similar philosophy on different fields of application can be found also in \cite{HAOCHENG2011}, 
\cite{zill2013}, and \cite{KORAMBATH2014}. Nevertheless it is worthwhile underlining that the objective
of \textit{LincoSim} is not providing  a framework to build a database system to enable conceptual design of a given 
engineering application. Instead \textit{LincoSim} is a platform which fully automatizes the complex work-flow of naval
simulation allowing a \textit{designer user} to perform and manage state-of-the-art 
3D Reynolds Averaged Navier-Stockes (RANS) CFD studies without the necessity of having CFD software 
competences nor having HPC skills to execute codes in feasible computing times.

In this regard, the first task towards LincoSim has been the standardization of data and operations useful
in the context of a common naval design cycle.

This work is structured as follows:
\begin{itemize}
    \item Section 2: is an exhaustive overview of \textit{LincoSim} components including a description of the web services and application components and of the computational tools
    \item Section 3: is actually an how-to set of subchapters that shows in a step-by-step way the possibilities available within the preconfigured workflow activities
    \item Section 4: is dedicated to the web application details including front-end, back-end and HPC interaction layers
    \item Section 5: discusses the CFD technical details involved in the implementation of the application solvers
    \item Section 6: presents conclusions and perspectives of the present work.
\end{itemize}

%% file: overview.tex
\section{Overview}\label{section:overview}
\begin{figure}
    \centering
    \includegraphics[width=\columnwidth]{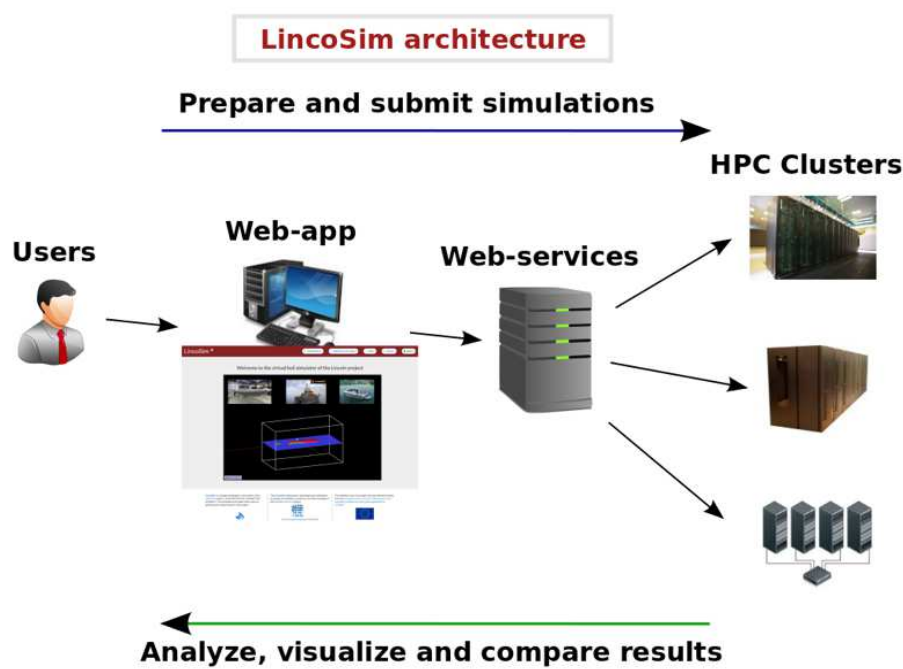}
    \caption{\textit{LincoSim} high-level workflow and architecture}
    \label{fig:overarch}   
\end{figure}

The high-level idea undergoing the development of the \textit{LincoSim} web application is to enable hull designers to simulate more and in advance to define more precisely reliable design solutions before moving to prototyping the hull and to do towing tank analysis and seakeeping only on a very limited set of well selected concept designs. 
To reach this point we collected feedbacks from naval companies and designers involved in the LINCOLN project in order to define an appropriate standardization of the input, output and of the computational tools aiming at configuring  a flexible automatic workflow template. 
This workflow is firstly based on a-priori knowledge of the physics of the problem and then tailored to perform well on the \textit{average} case of interest of the considered industrial cases. Different CFD templates are then obtainable customizing the default one to specific needs. 
The \textit{designer user} will be able to handle in just a few mouse clicks a relevant set of key-parameters of the hull hydrodynamics including: resistance curve, attitude, hull pressure distribution, waves distribution, wetted surface area and any other derived quantities. This is a set of KPI that matches the typical towing tank session. Nevertheless thanks to the \textit{LincoSim} application they are available within few hours and accessible via a web browser. The overall architecture is sketched in figure \ref{fig:overarch}.

In this section, we are going to briefly present the main features of the \textit{LincoSim} application. 
At the same time, we are going to detail the \textit{LincoSim} way to standardize metadata and tasks common in the context of a typical
naval design procedure.
In order to supply a solution to the requirements we design an automatic workflow divided into three main steps:
\begin{itemize}
    \item \textbf{pre-processing}: before starting any simulation, a valid geometry must be available. All the geometries must be uploaded using STL or OBJ file format (list of triangles). After this step, uploaded geometry can be visualized in interactive 3D view, moreover each uploaded geometry must be validated by the \textit{LincoSim} server validator which guarantees that the geometry file is suitable for the meshing and solving computations. If the user cannot successfully validate his geometry, an additional \textit{Request help} button is provided so that \textit{LincoSim} stuff can inspect and possibly solve the geometry problems. The database of owned and validated geometries is then available to the user so that he can easily access and use them for the simulations. In \textit{LincoSim}, the ownership of the entities (geometries and simulations) is shared among group of users which are called organizations. Organizations are managed by \textit{LincoSim} administrators following the requests of users.
    \item \textbf{CFD run definition}: in order to define a CFD run the user is requested to supply just a limited set of physical input values related to the desired performance of the hull CAD and none of these requires a specific CFD knowledge. Once a simulation has been defined, the user can review his choices by means of a synthetic table and  by means of the visualization of the 3D geometry including a 3D object that represents the velocity vector and the location of the Center of Gravity of the Hull and the free surface position. To start the simulation there are then two options: (i) simple-run: run the simulation as is (ii) multi-run: run many simulations according to a selectable physical parameter that varies over a range and with a given resolution. Typical range parameter will be the hull velocity in order to obtain a resistance curve. When clicking the \textit{Run simulation} button the simulation job is prepared and submitted to the specified HPC system. All the technical actions involved in this step are automatic and \textit{transparent} to the user.
    \item \textbf{CFD data analysis}: once a simulation is terminated, the user can analyze the outcomes of the single case present in the user database. For each simulation, in general, six tabs are available to categorize contents: simulation summary, hull dynamics time-series, wave elevations (2D), hull pressure distribution (3D), hull Pressure plots (1D), hull water-line (2D/1D). Comparative analysis and interactive 2D plots are also available for the user including parallel coordinates data clustering for fast and effective decision making.
\end{itemize}
Each step is designed in order to:
\begin{itemize}
    \item be able to start with just a limited set of input parameters (none of these related to CFD or HPC knowledge and defined by the \textit{designer user})
    \item allow for interactive visualization and analysis of inputs and results including 2D and 3D actions (zoom, sub-sample selection, rotating, parallel coordinate clustering and so on)
    \item be effective to the problem of hull hydrodynamics but also \textit{generic} allowing for further refinement and tuning for a specific case if required
    \item be robust on managing a wide range of possible inputs values.
\end{itemize}

From the technical point of view, the software stack of the \textit{LincoSim} GUI includes a front-end web interface based on Angular 5 JavaScript framework (\cite{angular}) for intuitive user interaction and embedded visualization libraries for 2D (plotly \cite{plotly} and d3js \cite{d3js}) and 3D (threejs, \cite{threejs}) plots. The back-end service implements several web-services using Python web2py framework (\cite{web2py}) served through NGINX reverse HTTP proxy.
Some tasks are performed asynchronously using the Celery (\cite{celery}) task queue. The database interaction is granted through Data Abstraction Layer pyDal (\cite{web2py}) currently connected to a PostgreSQL database (\cite{postgresql}), but modifying is straightforward. Search engine has been setup for advanced simulation search using an ElasticSearch (\cite{elasticsearch}) instance that replicates all simulation metadata of the database. HPC machines have no running \textit{LincoSim}-services: the web-services manage the HPC job submission and extract the data when a job is complete or the job status when it is running. Finally, the whole instance of \textit{LincoSim} platform is deployed using a Docker container (\cite{docker}).

The CFD undergoing engine is developed using the OpenFOAM toolbox enabling to solve the 3D Navier-Stokes equations including dynamic mesh motion and a number of Degrees of Freedom equal to $0$,$1$ or $2$ in a cost-effective way. The solution of hull hydrodynamics involves the solving procedure of a set of differential equations for a full 3D domain. The incompressible Navier Stokes for two phases (air/water) with interface tracking and capturing and the rigid body dynamics of the hull in equilibrium under the effect of gravity (hull mass), hydrodynamics forces (drag, lift) and momentum must be solved to get information about the hull hydrodynamics performances. All these equations are already included and available for direct usage in the OpenFOAM as standard solver (\textit{interDyMFoam}). This solver has been parameterized (for a set of key parameters) in order to build a generic template to solve the same set of equations on a generic hull geometry under a generic set of input parameters. The solution of hull hydrodynamics can take full advantage of the available HPC platform in many ways:
\begin{itemize}
    \item the availability of a large number of computational cores interconnected with high performance networks (low latency/large bandwidth) allows for split the computation of the single hull CFD on a set of computational cores (let’s say hundreds) and obtain a lower time to result;
    \item Several CFD runs can be performed simultaneously thanks to open-source license modelling of the selected solver. This kind of way of doing can also be performed using other third-party CFD software but requiring a potentially very large licensing budget.
\end{itemize}
Data of the hull dynamics are sampled for final review, moreover data at the final equilibrium status of the hull are processed in order to extract relevant KPI for the \textit{designer user}. OpenFOAM easily allows for data saving using VTK file format. All the data processing procedure is performed using Python programming language and dedicated modules: Numpy (\cite{numpy}) and VTK (\cite{pyvtk}).
Once data are processed, the publication on the \textit{LincoSim} web GUI is performed by means of semi-column tabulated files (csv).

%% file: usage.tex
\section{Usage workflow}\label{section:usage}
In this section, we are going to describe the \textit{LincoSim} application usage by means of a how-to set of subchapters that shows in a step-by-step way the possibilities available within the preconfigured workflow activities.

\subsection{General considerations}
\begin{figure}
    \centering
    \includegraphics[width=\columnwidth]{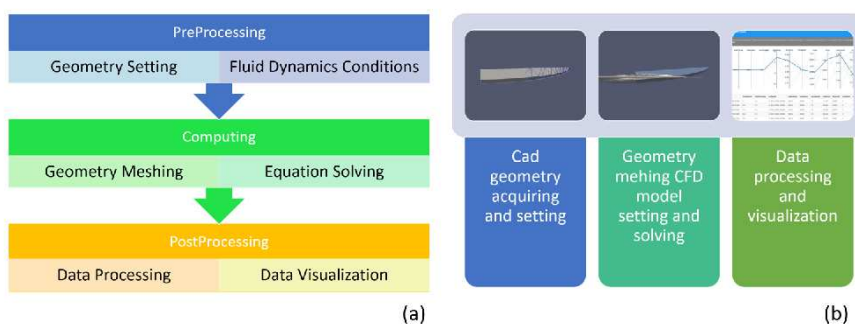}
    \caption{CFD hydrodynamics typical workflow: (a) boxes with steps; (b) boxes with sketches }
    \label{fig:overcfd}   
\end{figure}

In general, the \textit{LincoSim} platform allows to reproduce in an automatic way the CFD workflow for a given hull under a given working condition.
The CFD workflow for hull hydrodynamics problems is constituted by three main elements (see figure \ref{fig:overcfd}):
\begin{itemize}
    \item Inputs: made of two main parts that are the geometry description of the hull and the fluid dynamics conditions at which the hull has to be tested.
    \item Computing setup: that defines what kind of physics has to be solved and how. In other words the computing setup is the element of the workflow that defines what kind of \textit{experiment} we want to virtualize. In hull hydrodynamics study there are at least three types of test that can be performed: \textit{captive} or zero DoF, one DoF, two DoF. 
    \item Outputs: made of an arbitrary set of KPI values, tables, plots and diagrams that are necessary to support a ranking of the hull performances and support decision-making designer’s activity. 
\end{itemize}

It is worthwhile commenting a little bit more in detail the basic three computing setups that are available for hull hydrodynamics problems since these are the three main choices that a \textit{LincoSim} end-user can select during his working activity.
The zero-DoF (Degrees of Freedom) computational setup is representative of so-called \textit{captive} experimental condition in which the hull attitude is locked.
The one-DoF setup is used to represent a so-called \textit{free sink} experimental setup in which the hull is free to change his CoG position only by means of a rigid translation along the vertical axis.
The two-DoF setup is used to represent a so-called \textit{free sink and trim} experimental setup in which the hull is free to change his attitude by means of rotations around the transversal axis as centered in the Center of Gravity (CoG) and by means of a rigid translation along the vertical axis.
All these three setups are provided under calm water condition. However, in \textit{LincoSim} it is possible to specialize the computing template creating new simulation type: this is done configuring a so called \textit{simulation setup} so that users can then employ it in their runs.

\subsection{Geometry Input}
\begin{figure}
    \centering
    \includegraphics[width=\columnwidth]{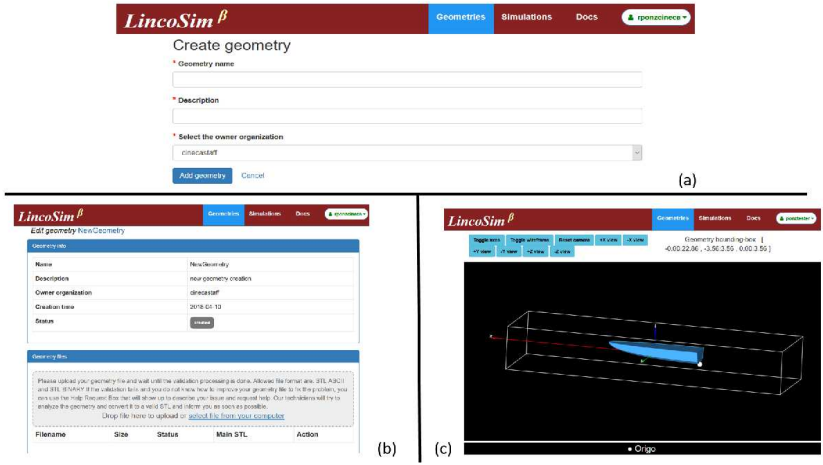}
    \caption{Geometry interface: (a) creation form; (b) edit form and file upload; (c) 3d interactive visualization.}
    \label{fig:geointerface}   
\end{figure}

Computational Fluid Dynamics (CFD) techniques require as a first step the import of CAD design shape (the hull). 
In \textit{LincoSim} the geometry model import is managed by means of dedicated section (geometry). 
In figure \ref{fig:geointerface} the geometry interface is provided.

The user is simply requested to select his input CAD file and upload it to the system. For sake of compatibility with the selected computational engine the file format accepted today is stereolithography or object (stl or obj file extension). These file formats are a very simple geometry representation made of a set of triangles and normals and are available as export file format form all the CAD design softwares. Once uploaded, the geometry file is processed, validated and displayed as 3D interactive object before being available for usage within the \textit{LincoSim} environment. Only validated geometries are available for usage. The used coordinate system is the absolute one and the positive (advancing) direction of the hull is considered as x-axis positive direction; the hull geometry must be prepared accordingly before uploaded.

Unfortunately, CAD design constraint and CFD ones are not the same. More important a CAD design that might looks correct in the CAD software can be considered as not valid once imported and used for CFD. Moreover, in our application we are moving from a real CAD file format (that depends on the used software) to a very simple triangulated file format.
Known critical CAD issues are differences in tolerance, presence of gaps due to misleading adjacent surfaces, triple connected edges, dirty construction edges or multiple normal definitions (outward or inward). 

Disregarding to what the CAD software is used there are some general rules that can be followed to achieve a good quality geometry for CFD, e.g. avoid self-intersecting surfaces, avoid small and much skewed patch-like surfaces added to join not adjacent surfaces, cleanup construction lines once they are not useful, cleanup construction points and in general eliminate multiple unnecessary points, align all the normals in a coherent way (all outward is the best).

Each CAD software has his tips to grant the best quality for exporting to stl. Moreover, since CFD works with closed volumes, if available, a closeness check within CAD software before exporting is recommended. Watertight geometry is mandatory to work with in CFD applications. If a water-tightness check is available in CAD software, an additional check that the geometry is a single volume instead of a set of interconnected faces or shells.

However, geometry preparation rules are not rigid and are strongly CAD software dependent. Therefore, the process of geometry preparation cannot be automatized and is left to \textit{LincoSim} user.

\subsection{Working conditions inputs}
Once the desired hull geometry is correctly uploaded and validated the user can start a virtual experiment using the desired hull geometry by using the \textit{Create simulation} button in the simulation page. 

There are two main sets of inputs that the user is requested to insert in order to start a new simulation: basic info and physical parameters.
\paragraph{Basic info}
        \begin{itemize}
            \item Simulation name: is a free entry and represents the name of the simulation in the user simulation database.
            \item Owner organization: is a drop-down menu where the user is free to select under which (if more than one) organization desire to perform the simulation.
            \item Simulation setup: is a drop-down menu so that only the simulation setups designed for a given organization are available for selection.
            \item Machine: is a drop down menu so that only running HPC infrastructures are available for selection.
            \item Geometry: is a drop down menu so that only the owned validated geometries are available for selection.
            \item Availability:  is a drop down menu and two kinds of value are allowed (public, private). Public means opened to all the registered users that have access to the \textit{LincoSim} platform. Private means private to the members of the organization under which the user is doing the simulation.
        \end{itemize}
\paragraph{Physical parameters}
        \begin{itemize}
            \item Hull mass: is a single scalar value that represents the total mass of the hull expressed in SI units.
            \item Hull Center of Gravity: is a tuple of three values that represents the coordinates of the center of gravity (CoG) of the hull in the absolute reference system.
            \item Hull velocity: is a single scalar value that represents the velocity magnitude of the hull in SI units.
            \item Water temperature: is a single scalar value that represents the temperature of the water in SI units.
            \item Hull inertia: is a tuple of three values that represents the value on the diagonal of the matrix of moments of inertia (technically, second moment of area) of the hull geometry computed for the three main axes (xyz) respect to the CoG coordinates expressed in SI units.
            \item Water z-pos: is a single scalar value that represents the starting z coordinate of the waterline in the absolute reference system.
            \item Wave height: is a single scalar value that represents the guessed maximum height of the wave in the absolute reference system.
            \item Hull trim angle: is a single scalar value that represents the starting trim angle of the hull in the absolute reference system computed for a rotation around the CoG.
        \end{itemize}
Notably none of these values is related to any CFD or HPC knowledge, instead all the inputs parameters are strictly related to well-known hull design parameters that can be computed easily with common CAD design software.

\subsection{Complete Single and Multiple Analyses}

Starting from the geometry and the working conditions, it is possible to start
a simulation by clicking on the \textit{Submit simulation} button.
This kind of application represents the average case for which the \textit{LincoSim} 
application has been designed. 

Moreover, to support common parametric studies we designed an additional submission type, called \textit{Range simulation}. This simulation type allows to quickly and effectively submit a set of simulations in which one of the physical input parameters is free to change in a range defined by the user whereas all other inputs parameters remain constant. The typical example of this kind of analysis is the so-called \textit{resistance curve} analysis in which the designer needs to get an understanding of the total drag value of the hull at different velocity conditions keeping all other physical parameters fixed. Another meaningful example can be a set of captive cases with different trim angles. 

It is worthwhile stressing here that the availability of standardized, automatically coherently streamed multiple case workflow is a strong benefit of \textit{LincoSim}.

\subsection{Outputs}
\begin{figure*}
    \centering
    \includegraphics[width=\textwidth]{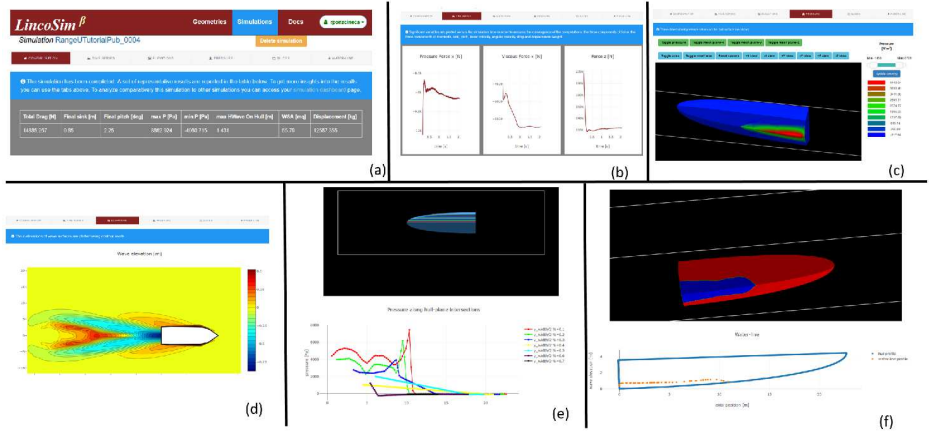}
    \caption{Example snapshots of result table and interactive visualizations: (a) summary table; (b) time-serie plots; (c) hull pressure 3d view; (d) wave elevation plot; (e) pressure along hull slices; (f) wetted surface over hull: 3d and 2d plots.}
    \label{fig:results}   
\end{figure*}

For each simulation, there are two main kinds of output automatically provided by \textit{LincoSim}: 
\begin{itemize}
    \item Synthetic key parameter index (KPI): are analytical values that are typically available as outcomes to the designer in order to rank the hull performance. In particular, \textit{LincoSim} autonomously compute these main hydrodynamics quantities of interest: total drag, maximum/minimum pressure value on the hull, maximum wave height on the hull, wetted surface area of the hull (wsa). An example of summary table is provided in figure \ref{fig:results}.(a).
    \item Visual data: are 1D, 1D over time, 2D and 3D datasets that are interactively available to the end user to get also a visual insight on a wide range of quantitative outcomes of the performed analysis. Deeper analysis can be performed looking at: pressure patterns on hull, wave patterns, pressure patterns over selected longitudinal lines on hull, forces acting on hull time history, CoG dynamics time history, wetted surface distribution on hull. Example visualizations of these interactive visualizations are shown in figures \ref{fig:results}.(b)-(f).
\end{itemize}

With positive completion of the analysis, the user is allowed to access KPI, visualize, and interactively navigate processed data.
Nevertheless, in order to support any kind of error (system error, modelling error, solver error, data processing error or during cad import error), \textit{support request} boxes have been designed and added to the \textit{LincoSim} application for both geometry and simulation sections.
Thanks to the two help request boxes, the end-user experiencing a problem during his normal workflow stream can easily inform \textit{LincoSim} personnel to give support for the specific given problem. An automatic mailing system contacts the technical support.

\subsection{CFD setup customization}
As introduced, the automatic CFD workflow is usually made of three main parts: meshing, computing, post-processing. Changes to one or more of these main blocks identify a different workflow type. A general single approach valid for every kind of hull under any kind of working condition is not feasible and it would lead to a very monolithic and rigid approach that from our point of view would be highly inefficient. For this reason, we decided to design our application starting from macro settings, reflecting specific general needs, which can be further customizable upon request.

For instance the three main kinds of workflows available for each end-user are:  the ``captive'', the ``1DoF'' and the ``2DoF'' as explained.
These three workflows are already designed to be able to work efficiently on specific end-user types of hull. 
Nevertheless, the three main blocks can be, to some extent, also customized and fine tuned to reach some specific needs. \textit{LincoSim} admins can configure additional so called \textit{simulation setups} which work automatically and are presented to the end-users as black-boxes. This approach allows for maximum flexibility. Once the user identifies a specific need and desire to request a customization of the workflow he can provide technical details, including benchmarked reference cases, to LinsoSim developers and request the desired customized version of an available workflow. This clearly means that an external intervention is needed but, on the other hand, \textit{LincoSim} is ready to be extended.

\subsection{Geometry and Simulation Dashboards}
\begin{figure}
    \centering
    \includegraphics[width=\columnwidth]{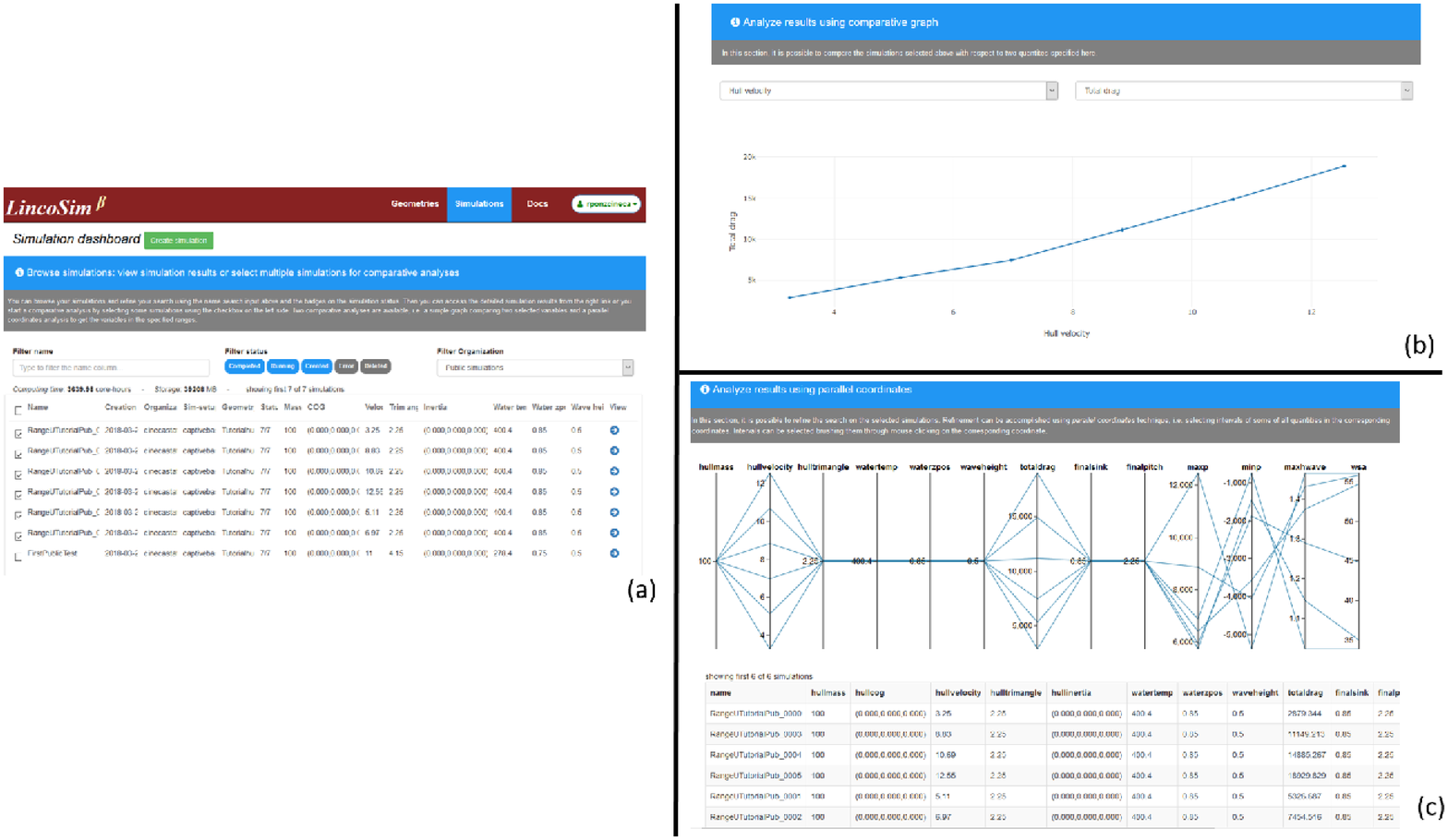}
    \caption{Simulation analysis tools: (a) searchable dashboard; (b) variable comparison plot; (c) parallel coordinates plot with refinement table.}
    \label{fig:simdash}   
\end{figure}

Management of geometry and simulation database is another of the key points within the \textit{LincoSim} application. Thanks to the unified and standardized approach adopted, users can visualize in a very friendly and clean way two smart tables displaying the different geometries uploaded to the system and the different simulations performed, respectively (see figure \ref{fig:simdash}.(a) for an example of a simulation dashboard). 

Moreover, due to the intrinsic richness of the dataset, the simulation dashboard has some specific additional features.
In particular, there are several filters and analysis tools available:
\begin{itemize}
    \item Filter by name: a standard by name filter is available to sort out only desired simulation sets.
    \item Filter by simulation status: quick tags selection concerning the actual status of the simulation can be easily performed. The available statuses are: Completed, Running, Created, Error, Deleted
    \item Filter by owner organization: a user that performs simulations for different organizations can easily pick up only the subset of data related to a selected organization that has been declared as owner during the submission process
    \item Analyze results using comparative (see figure \ref{fig:simdash}.(b)) graph for a set of selected hulls: using this feature it is possible to compare the simulations selected above with respect to two quantities specified here.
    \item Analyze results using parallel coordinates (see figure \ref{fig:simdash}.(c)) for a set of selected hulls: using this feature it is possible to refine the search on the selected simulations. Refinement can be accomplished using parallel coordinates technique, i.e. selecting intervals of some of all quantities in the corresponding coordinates. Intervals can be selected brushing them through mouse clicking on the corresponding coordinates.
\end{itemize}

%% file: web_details.tex
\section{Web services and application Implementation}

\subsection{From requirements to technologies}

The overall architecture design of \textit{LincoSim} and the choice of technologies have been done with particular care considering the collected requisites. 

First, we considered existing frameworks which try to ease the HPC access barrier to run applications.
A number of tools is available nowadays, e.g.
\begin{itemize}
	\item 
		OnDemand (\cite{ondemand}): defined as ``one stop shop for access to our High Performance Computing resources. With OnDemand, you can upload and download files, create, edit, submit, and monitor jobs, run GUI applications, and connect via SSH, all via a web broswer, with no client software to install and configure''
	\item EnginFrame (\cite{enginframe}): defined as ``advanced, commercially supported HPC Portal in the industry, with a proven track record of successful production deployments within corporate networks and research clusters. EnginFrame enables efficient Inter-Intranet access to HPC-enabled infrastructures. HPC clusters, data, licenses, batch \& interactive applications can be accessed by any client using a standard browser. The open and evolutionary framework of EnginFrame is based on Java, XML and Web Services, and facilitates deployment of user-friendly, application- and data-oriented portals''
\end{itemize}
While a comprehensive comparative analysis of these tools is beyond the scope of this document, it turns out that these tools aim at minimizing the user effort required to deal with HPC resources: queue system commands, software installation, main application input parameters are already set so that the user can focus on its main objectives and results. To achieve this, these platforms include a set of middleware components which can be used as building blocks to prepare attractive Graphical User Interface (GUI) linked to easily configurable HPC computations. The spread of these tools proves their benefit in the HPC workflows. 
However, EnginFrame-like platforms are not designed to manage full automation of the workflows even if it is somehow possible to manually implement such work-flows strongly customizing the underlying code. In any case, these platforms still require the user to master advanced software, often manually integrate pre/post-processing stages, deal with files, and so forth. 
For the \textit{LincoSim} platform the capabilities of these tools are not enough to reach its goals. 

We decided to move to a more high-level approach where the user is not supposed to be HPC expert at all -- it can even ignore which cluster he is using and any detail on the queue system -- and also not a CFD expert at all -- it can even ignore which application codes are used to simulate its hull dynamics. The \textit{designer user} has only to be expert on the geometry design and physical parameters (input) and on the analysis of the results (output). 

This \textit{black-box} user perspective allows to achieve the most critical \textit{LincoSim} requirements, in particular:
\begin{itemize}
	\item Suitability for non expert computational scientist: the platform should provide a minimal set of inputs and allow to execute the simulation automatically; it cannot be completely automatic but the choice should be made as simple as possible
	\item Transparency-on-computing-machine: the platform should be able to launch the computations on different machines/clusters as transparent as possible for the user.
\end{itemize}
    Clearly, such high-level platform requires a huge implementation effort to be prepared. 
To select the best technologies for \textit{LincoSim} we considered two other significant requirements (not managed by EnginFrame-like tools as well):
\begin{itemize}
	\item Simulation management: not only the platform has to allow to run simulations but it should allow to organize and compare contents through intuitive dashboards
	\item Groups management: there must be the possibility to keep geometry and simulation data private and there should be the possibility to share simulations across specific set of users.
\end{itemize}
Such tailored requirements suggest to move to more customizable technologies where the main architecture is freely designed from scratch, i.e. we decided to write a complete application based on state-of-the-art libraries and tools.

\subsection{Components and software stack}

From the technical point of view, the description of the software stack of \textit{LincoSim} follows:
\begin{itemize}
	\item  front-end:
		\begin{itemize}
			\item is based on Angular 5 JavaScript framework  (\cite{angular}) for intuitive user interaction and dynamic content retrieval. 
			\item advanced visualizations are possible through embedded libraries:
				\begin{itemize}
					\item three.js (\cite{threejs}) for 3D WebGL  (\cite{webgl}) visualizations, e.g. meshes and fields;
					\item plotly.js (\cite{plotly}) for 1D visualizations, e.g. temporal series, and 2D, e.g. contour plots;
					\item D3.js (\cite{d3js}) for parallel coordinates analyses;
				\end{itemize}
		\end{itemize}
	\item back-end:
		\begin{itemize}
			\item is served using HTTP reverse proxy server provided by NGINX (\cite{nginx})
			\item web-services are provided through web2py Python framework (\cite{web2py})
			\item asynchronous tasks are performed by Celery task queue (\cite{celery}) on top of RabbitMQ message broker (\cite{rabbitmq})
		\end{itemize}
	\item metadata management
		\begin{itemize}
			\item the database interaction is granted through Data Abstraction Layer pyDal currently connected PostgreSQL database (\cite{postgresql}). 
			\item a search-engine has been setup for advanced simulation search and filtering using an ElasticSearch (\cite{elasticsearch}) instance that replicates all simulation metadata of the database
		\end{itemize}
	\item HPC machines
		\begin{itemize}
			\item have no running LincoSim-services: the web-services manage the HPC job submission through the message queue and the HPC queue and extract the data when a job is complete or the job status when it is running. 
			\item the web-services directly access HPC storage to visualize results
			\item currently PBS (\cite{pbs}) and SLURM (\cite{slurm}) workload managers are supported.
		\end{itemize}
\end{itemize}

Web-server can be hosted on a common Virtual Machine and has been tested on a OpenStack (\cite{openstack}) virtual machine using 4 cores and 12GB Ram. The platform deploy is very easy using a preconfigured Docker (\cite{docker}) container.
    HPC machines can be connected to the \textit{LincoSim} platform with a limited effort: currently the platform has been tested connected to two Intel based clusters with different queue systems (PBS and SLURM).

\begin{figure*}[!h]
	\centering
	\includegraphics[width=0.8\textwidth]{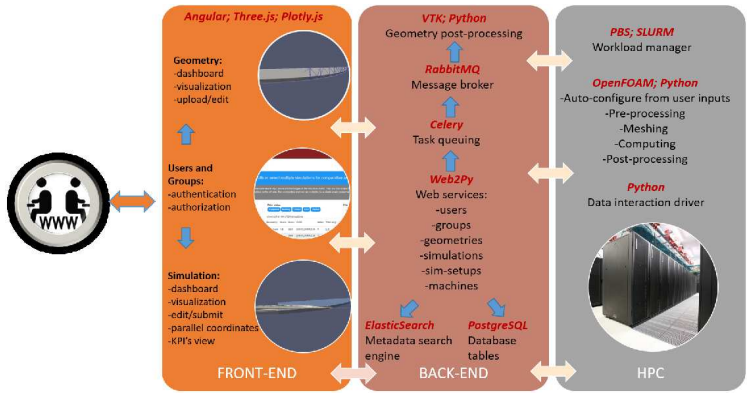}
	\caption{Sketch of user workflow including the main used tools with their roles}
	\label{fig:webtechuserwf}   
\end{figure*}

\subsection{User workflow}

    The user workflow has been already described in the section \ref{section:usage}. In figure \ref{fig:webtechuserwf} the main components of the platform with respect to user workflow are sketched with a summary of their main roles. Some implementation notes follow. 

We stress that the browser is the unique access point for the user. No direct interaction with web-server and HPC machines is needed. The single exception -- not represented in the figure -- is given by the mail that the HPC machines send to the user to inform him that the simulation has started or concluded.
    Let us describe what happens under the hood when the user performs his most important actions. After logging on, the user can access the two main platform parts: geometry and simulation management.

\paragraph{Geometry}
The geometry section interacts with the authentication and geometry back-end web-services to create/edit geometries, upload files, retrieve content and decimate the CAD geometries for visualization purposes. The database stores the metadata while geometry files are stored using the web-server storage so that they can be later used by any HPC machine which will perform the computation using that geometry. Since the decimation process -- which also acts as a validation -- may require a certain execution time, it is executed using an asynchronous task thus avoiding time delay when serving the web-page. 
\paragraph{Simulation}
The simulation section interacts with the major part of the web-services for several tasks (authentication, getting available machines, getting available simulation setups, creating simulation, etc.). In addition to the database and search-engine operations, some web-services require the connection to the HPC machines. For robustness reasons, the logic has been devised so that the interaction with the HPC machines is minimized, e.g. the simulation creation does not require a connection to the HPC machines. The simulation submission requires the connection but this is performed through the asynchronous task queue which repeats the submission attempts until the selected HPC machine is capable of receiving it. After the submission, the simulation job starts doing its tasks and -- step-by-step -- calls \textit{LincoSim} services to update the status on the database so that the user can check it on the webpage. The user is also informed by mails when the job starts or ends. When the job is completed the user can access the table of results as well as visualize the results. The access to results clearly requires the connection of the web-server to the HPC machine.

    \subsection{Admin workflow}

\begin{figure*}[!h]
	\centering
	\includegraphics[width=0.8\textwidth]{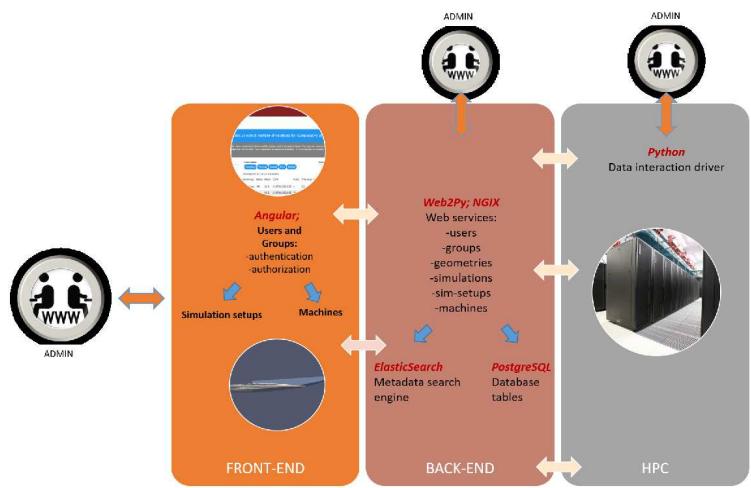}
	\caption{Sketch of admin workflow including the main used tools with their roles.}
	\label{fig:webtechadminwf}   
\end{figure*}

    The admin workflow is summarized in figure \ref{fig:webtechadminwf}. In contrast to what happens to user workflow, skills as system administrator, HPC specialist and CFD engineer are required. The admin usage has three access points:
\begin{itemize}
	\item web access to admin panels:
		\begin{itemize}
			\item organizations panel: from this panel it is possible to define groups of users which share authorization privileges on machine and simsetup usage
            \item machines panel: from this panel it is possible to configure the connection of a HPC machine to be used by \textit{LincoSim} simulations
            \item sim-setups panel: from this panel it is possible to configure a simulation setup which will be used by \textit{LincoSim} simulations
		\end{itemize}
    \item ssh access to web-server machine: access to web-server machine is required to install \textit{LincoSim} front-end and back-end components. The usual admin tasks involve:
		\begin{itemize}
			\item the management of services 
			\item the direct access to logs and in particular error logs
		\end{itemize}
	\item ssh access to HPC machines:
		\begin{itemize}
			\item machine configuration: a set of predefined files and folder must be prepared in the HPC machine to be connected alongside the needed software stack (mainly Python libraries and the CFD solvers which will be used)
            \item sim-setup configuration: whenever a simulation setup is added to the system, the corresponding solver code must be prepared according to the \textit{LincoSim} rules
		\end{itemize}
\end{itemize}

    It is expected that admin staff interacts with users at least when experiencing errors but also in other circumstances as, for instance, to get feedback on the quality of results so that admins can have hints on how to improve a simulation setup workflow.

\subsection{Users and groups}

    The \textit{LincoSim} management of users allows to restrict permissions on the main components of the platform. Permissions are enforced at front-end and back-end level for clarity and security reasons. The web-server authorization mechanism of \textit{LincoSim} is implemented by the Role Based Access Control (RBAC) provided by web2py. Permissions are group-based, i.e. privileges are granted to \textit{LincoSim} organizations that are groups of users managed by \textit{LincoSim} admin via the \textit{admin organization panel}. 

    Each member of a \textit{LincoSim} organization:
\begin{itemize}
	\item can create and manage geometries of its organization(s)
	\item can create and manage simulations of its organization(s)
\end{itemize}
    This means that geometries and simulations are shared among users of the same organizations. Since a user can belong to many organizations, the sharing possibilities are wide.
    Each organization, according to the privileges granted by \textit{LincoSim} admin, can:
\begin{itemize}
	\item submit simulations to the authorized subset of computing machines
	\item submit simulations using the authorized subset of simulation setups
\end{itemize}
    The restricted access to computing machines and simulation setups is not only a privilege restriction but also a way to ease the usage since each organization should be configured so that only the needed sim-setups and machines are available.
    When creating a simulation it is possible to select a visibility option which can be \textit{private} or \textit{public}. Private simulations can be accessed only by members of the owner organization while public simulations can be accessed by any logged user. 
    User registration is subject to approval by \textit{LincoSim} administrators.

\subsection{Simulation setups}

    A simulation setup contains the logic of computational solver including pre and post-processing stages. \textit{LincoSim} allows to configure completely different simulation setups provided a number of constraints on input and output quantities are fulfilled. To add a new simulation setup, \textit{LincoSim} admin has to create a setup from the web interface. 
    In addition to sim-setup name, default parameters and simulation statuses dictionaries must be provided. Also, the HPC machines supporting this setup have to be specified
    After this step, the \textit{LincoSim} admin has to access the HPC machines where the setup is defined and prepare a folder with the simulation setup logic. The sim-setup logic contains all the commands, libraries needed to perform the simulation. The starting file is the \textit{build script}, i.e. a Python code which writes the ``simulation'' job according to the metadata information of a simulation (extracted from \textit{lincosim.inp} metadata starting file).

\subsection{Front-end implementation}

    The front-end design of \textit{LincoSim} has a key role with respect to the overall usability of the platform. In particular since the entire user experience is web-based, powerful web frameworks and libraries have been selected:
\begin{itemize}
	\item JavaScript framework: modern frameworks allow to manage huge projects in a well-organized and efficient fashion. We used the state-of-the-art Angular framework. The implementation of strongly interactive web-content (forms, tables, figures) is a typical scenario suitable for Angular project. In particular the simulation dashboard page allows a large number of actions executable by the users and synchronization with the charts. The Angular adoption also significantly improves the maintainability of the source code.
    \item Visualization libraries: the implementation of embedded visualization tools in the \textit{LincoSim} web-application allows to easily handle dynamic content avoiding the need for external general-purpose visualization softwares dramatically simplifying the overall usability of \textit{LincoSim} itself.
		\begin{itemize}
            \item three.js a popular cross-browser JavaScript library used to create and display attractive 3D computer graphics in a web browser. Three.js uses WebGL to get good performances. The library has been intensively used in the front-end core of \textit{LincoSim} to display interactive -- rotation, zoom, pan -- views of the relevant geometries and fields (pressure, wet surface). STL and VTK formats are used as interchange formats: STL for geometry and VTK for the resulting field. Even if three.js comes with a wide number of readers and options, significant work has been needed to adapt it to the type of visualizations more useful in the \textit{LincoSim} context. The quality of result is comparable to the usual results in external visualization software. Having anything integrated avoids the effort by the user to interact with files, install software and know what and how to visualize things.
			\item plotly.js, built on top of d3.js (\cite{d3js}) and stack.gl, is a high-level, declarative charting library including  20 chart types, 3D charts, statistical graphs, and SVG maps. We use it for 2D line plots and elevation plots. In addition to the attractive results, the possibility to zoom and export images are a significant plus.
			\item d3.js is used to implement the parallel coordinates tools available on the simulation dashboard.
		\end{itemize}
\end{itemize}

\subsection{Back-end implementation}

\paragraph{Web-services}

    Web-server architecture employs a first reverse proxy layer provided by NGINX which allows to conveniently handle the different types of requests. Besides, we adopted the web2py Python framework, a free open source full-stack framework for rapid development of fast, scalable, secure and portable database-driven web-based applications. Among other things, web2py is a convenient choice due to the native integration with the database component (through the PyDal data abstraction layer), to a solid error handling, and a clear admin dashboard. According to the typical design of an Angular application, we decided to implement web-server controllers as web-services API, i.e. functions which exchange model data (mostly using JSON format) with no direct reference to the template view logic. The two main advantages of this approach are:
\begin{itemize}
	\item clear separation among view logic -- using Angular -- and back-end logic -- using web2py web-services
	\item possibility to execute any web-server task via command line or by other interfaces calling APIs
\end{itemize}
    Web-services are organized in 6 controllers -- auth, organizations, simulations, geometries, machines, simsetups -- with a total amount of about 70 API.
    The uwsgi (\cite{uwsgi}) module is used to manage the web2py Python layer from NGINX.

\paragraph{Database and metadata catalogue}

    The database is an important component of \textit{LincoSim} since it allows to organize the different entities keeping track of the evolution of the simulations and geometries of the user groups. The structure of database is quite simple (10 tables). Even if at the first stage the number of database entries will be limited, we decided to connect a full PostgreSQL database to potentially handle intensive usage. The database interaction is performed by web-services -- and by Celery tasks invoked by web-services -- through pyDAL, the pure Python data abstraction layer provided by web2py, thus providing a clean and maintainable source code.
    A special table of the database is the simulation table. Since the simulation dashboard allows advanced search and filter actions, we complemented the database with ElasticSearch, a specific and well-established search engine which easily allows to perform all kinds of search and filter \textit{LincoSim} has to propose to the user. Data of simulation table are directly replicated to the ElasticSearch instance. The Python API for ElasticSearch allows to easily interact with the service from the web2py services.

\paragraph{Asynchronous tasks}

    Two web-services perform critical tasks which is convenient to execute asynchronously:
    \begin{itemize}
	    \item  geometry validation and decimation, because of the significant running time
	    \item  simulation submission, because the submission implies the connection to the HPC machine which can be under maintenance (the task is repeated till success in that case)
    \end{itemize}
    Asynchronous tasks are managed by the solid Celery task queue system based on the RabbitMQ (\cite{rabbitmq}) message broker. Celery tasks are called by web2py web-services thanks to a proper interface layer. We also setup a Flower (\cite{flower}) instance -- an additional Celery monitoring tool -- which allows \textit{LincoSim} admin to keep track of the submitted jobs from an intuitive dashboard, useful especially in error conditions. 

\subsection{HPC interaction}

\paragraph{Adding machines}

    The design of \textit{LincoSim} allows to add and manage different external HPC machines to perform the simulations. Currently Linux based HPC machines with PBS or SLURM workload managers are supported, but future extensions are possible. To add an HPC machine, \textit{LincoSim} admin can first use an intuitive web-form

    where the basic information of the machine -- e.g., ip address, root folder, user names -- have to be inserted. After that, admin has to directly access the HPC machine and create a predefined \textit{LincoSim} tree structure with some predefined files. Such files contain the common queue submission and interchange logic. These files are called by the web-server code to start the simulation submission. 

\paragraph{Data interaction via Python Paramiko}

    The simulation data reside on the HPC machine where the simulation has been executed and must be accessed by the web-service whenever the user request a visualization page on the simulation. This is achieved using Paramiko (\cite{paramiko}) Python library. The library allows to remotely open the needed file and read it so that the web-service can then return it to the front-end interface. The same library also provides ssh-like Python access and is used to connect the celery submission task called by web-services to the HPC machine. 

\paragraph{Submitting and checking the statuses of jobs}

    The simulation submission follows four steps:
\begin{itemize}
	\item clicking on the web-page submit button, the \textit{submit} web-service API is called and it triggers the Celery async task \textit{async\_submit}
	\item \textit{async\_submit} Celery task connects to the HPC machines
		\begin{itemize}
			\item writes a \textit{lincosim.inp} JSON file which summarizes all the database information of the simulation
			\item submits to HPC queue system (PBS or SLURM) a \textit{prepar}  job
		\end{itemize}
	\item the \textit{prepare} job 
		\begin{itemize}
			\item reads and parses the \textit{lincosim.inp} file
			\item calls the sim-setup \textit{build\_prepare} script
			\item submit the \textit{simulate} job
		\end{itemize}
	\item the \textit{simulate} job performs the full CFD simulation -- including pre and post-processing -- following a step-by-step logic with a strict error check.
\end{itemize}
    Both the \textit{prepare} and \textit{simulate} jobs call back the \textit{LincoSim} web-services to update the status of the job so that the user can know the job status from the web interface. \textit{simulate} job also sends mail to the user when starting or ending the job.

\subsection{Error-handling}

    Given the overall complexity of the \textit{LincoSim} architecture, the large number of mutually interacting components -- most of which intentionally hidden to the user -- and the \textit{a priori} unpredictable modeled physics, the possibility to encounter an error condition cannot be excluded. Indeed, it is crucial that the user and administrative errors are explicitly treated -- especially from user side where no direct access to servers is possible.
    The most important sources of failure in \textit{LincoSim} are expected to be:
\begin{itemize}
	\item system failure: both Virtual Machines hosting the web-server and HPC clusters computing CFD numerical solutions are subject to diverse source of failure, e.g. file-system, connection, queue daemons, virtualization infrastructure
	\item  CFD algorithm failure: since the user can insert in the web form the physical values of its simulation as well as the CAD geometry, the CFD algorithm is expected to work correctly only when these values are physically reasonable. It is very difficult to know how to select reasonable ranges of parameters to be inserted. 
\end{itemize}
    \textit{LincoSim} manages the error handling through different approaches summarized below:
\begin{itemize}
    \item Graceful messages to users: web-services use standard \textit{try/except} strategy for critical sections while typical \textit{internal server errors} automatically save error ticket logs the web interface always providing clear messages to the users
	\item  Robust submission: simulation submissions is managed at web-server layer through a task queue which attempts to submit the job to the HPC queue until success; this strategy hides the down-times expected in a usual HPC environment
	\item  Statuses: the CFD computations follow a sequence of computing steps. For each step an error condition is signaled by a negative number with an error message which are reported to the end-user which can try to verify if the given geometry and parameters are physically reasonable. The geometry management also follows a status logic with a single step (validation successful or failed)
    \item  Mail to engineers box: for both geometries and simulations (which are in error conditions or not) it is possible, from within the corresponding page, to fill a textarea and press a button to send a message to \textit{LincoSim} engineers which can try to understand why geometry validation or simulation has failed
    \item Monitoring:  \textit{LincoSim} platform activity is monitored through cron script
    \item Database: given the complexity of interaction between web-server and HPC machines, an additional script verifies the consistency of the database
\end{itemize}

\subsection{Source code, container, and documentation tools}

    Given the high number of \textit{LincoSim} custom components and services, the installation, production management and maintainability of the code are critical. The usage of advanced tools to tackle these issues is crucial to obtain optimal results. We decided to adopt \textit{gitlab} to handle source code versioning  and manage development issues.

Platform deploy is achieved via Docker. The usage of a Dockerfile dramatically simplifies the installation effort required to prepare the code and launch the services. Docker containerization is more and more an enterprise adopted solution to provide a stable, replicable and adaptable environment for a wide range of components. In \textit{LincoSim} we decided to use a Docker container to run all the back-end and front-end services. The combined usage of Docker and gitlab repository allows to instantiate \textit{LincoSim} in a few minutes -- compared to the hours required to manually install of the components. Moreover, it can be installed always the same way and on hosts running different Operating Systems. We decided to use a single Docker container running all the required services and to adopt a process manager -- supervisor (\cite{supervisor}) -- to manage the processes. Currently, 8 services are managed by supervisor, i.e.: (1) celery; (2) elasticsearch ; (3) flower ; (4) nginx ; (5) postgresql ; (6) rabbitmq ; (7) sshd ; (8) uwsgi\_emperor.

The CFD solver code (OpenFOAM and simulation setups sources) are not deployed using Docker because of the intrinsic difficulties arising when using Docker in HPC contexts. However, given the modular design of \textit{LincoSim} that components can be adapted to the needs of the user without modifying the whole design of the platform.

Documentation of \textit{LincoSim} is crucial to provide success to the platform.  There are several types of online documentation, directly available from within the \textit{LincoSim} web page, to support the platform comprehension and usage:
		\begin{enumerate}
			\item In-page minimal usage instructions: by means of two fix textual boxes (blue box synthetic content and  grey box for more exhaustive informations) the user is informed about the operations that can be performed at that page.
			\item In-line description of input form fields: each free input box is equipped with a short inline description of how the input should be provided.
			\item Overlay info box opens on help button click: for some specific technical input that requires a specific convention knowledge an overlay information box is provided by means of a clickable question mark blue icon.
            \item User manual: for exhaustive description of each \textit{LincoSim} functionality a complete user manual is available.
				The user manual is provided through the Sphinx documentation generator so that impressive web and \LaTeX docs are easily avialable.
			\item Video tutorials: to support first time users a set of video tutorials is available. 
		\end{enumerate}

Finally, we decided to base our code on open-source libraries for both web-application and computational layer softwares.

%% file: cfd_details.tex
\section{Computational Modelling Implementation and Validation}
The computational core of the \textit{LincoSim} application is a CFD solver necessary to properly evaluate the set of equations involved on the physics of hull hydrodynamics. As previously discussed, many CFD workflows can be configured by \textit{LincoSim} admins and saved as so-called \textit{simulation setups}. These simulation setups can even employ different CFD engines but are always made available to end-users as block-boxes.

Currently the tested CFD setups are based on a standard de facto open-source library (OpenFOAM) based on the so-called finite volume method (FV). Finite volume method in general and the OpenFOAM toolbox in particular have been extensively proven to be reliable and efficient to model and solve complex 3D hull hydrodynamics problems. The basic idea undergoing the implementation in \textit{LincoSim} is to provide:
\begin{itemize}
	\item CFD solver selection
	\item Solver configuration and validation for a case of interest to prove that the physics of the problem is well caught.
	\item Instruct the validated setup in a parametric way in order to be able to autonomously build and solve CFD models for different input datasets.
\end{itemize}

\subsection{CFD Solver selection}
The physics of the problem that we want to face is quite complex since we must solve the 3D Navier-Stokes system of equations for a system made of two immiscible incompressible Newtonian fluids (air and water) under turbulent flow regimen with interface surface capturing including dynamic mesh motion and interface mesh morphing. Luckily, OpenFOAM is shipped with a considerable set of pre-configured solvers written to solve specific multi-physics problems, like the one involved by hull hydrodynamics, allowing the user to just select very specific details about the meshing strategy, the numerical settings, the solving strategy, the turbulent modelling and the data sampling. All these details selection is made by means of a set of input files organized as \textit{dictionaries} in which at each key parameter can correspond a wide range of possible values coherently to what the OpenFOAM toolbox contains. The name, number and type of dictionaries necessary to run a specific OpenFOAM solver depend on the specific solver.

The solver identified in our application within the OpenFOAM toolbox (version 5.0) list of available solvers is the so-called \textit{interDyMFoam}. 

\subsection{Solver configuration and validation}
In order to use the solver we must provide meaningful inputs to all the dictionaries involved in the use of the selected solver. Notably there are a set of dictionaries that are strictly related to the specific hull selected while others are more general and their content setting can be considered valid for a wide range of problems. During our development procedure, we therefore firstly worked on these general settings to find a good balance between solver robustness, accuracy and time to result. The validation procedure has therefore been performed to obtain a first general calibration of the solver to face the planing hull hydrodynamics.
To do so we selected an experimental validated reference and we replicated the same conditions using the OpenFOAM toolbox. The reference measurement campaign is taken from (\cite{begovic2012}). 
In that work a set of hulls has been extensively studied in a towing tank experimental facility at different velocities monitoring a wide set of KPI. A full validation campaign, which is undergoing, is beyond the scope of this work. 
Here we took one hull to design a proper automatic workflow and we used another one to test the workflow performances and compare the numerical results for a given set of KPI: total resistance, wetted surface area and pitch angle.

Using our solution strategy, we were able to obtain smooth solution patterns in a reasonable time 
For instance our reference case dynamics can be studied in about 6-10 hours.

\begin{figure}
    \centering
    \includegraphics[width=\columnwidth]{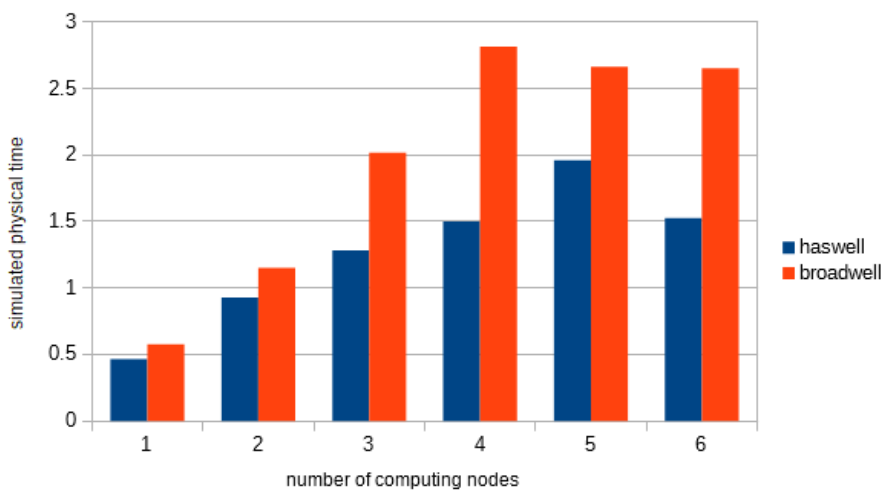}
    \caption{Scalability analysis: simulated physical time versus number of computing nodes for Cineca Intel Haswell cluster vs Intel Broadwell cluster for typical \textit{LincoSim} simulation}
    \label{fig:cfdscalab}   
\end{figure}

\paragraph{Parallel computing} 
A reasonable size for the computing mesh clearly depends on the hull size and grid refinement to be obtained. Besides, different simulation setups can be tuned according to the desired mesh refinement. From validated cases, we found that mesh ranging from 3 to 10 millions of cells are cost-effective choices. The usage of parallel computing on HPC clusters is crucial in order to get results quickly especially when parametric studies are needed (several simultaneous simulations) thus requiring significant intermittent hardware needs. Parallelization is well implemented in the vast majority of CFD software, and, in particular, in OpenFOAM. The scalability behaviour of a typical -- 5 millions of cells -- \textit{LincoSim} simulation using two common HPC architectures -- based on Intel Haswell and Broadwell CPUs, respectively -- is provided in figure \ref{fig:cfdscalab}. From that analysis, we decided to configure corresponding simulation setups to run in parallel over 3 full nodes so that speed-up efficiency is high ($91\%$ and $116\%$ respectively). 

\paragraph{Validation results} 
The quantitative KPI outcomes 
represents a synthetic integral evaluation of the distance between the numerical solution and the experiments.

The percentage difference between computed and measured values depends on the observed quantity thus reflecting how the numerical model parameters are sensible to our solution strategy. For the purpose of our solver setup these differences are considered within an acceptable bound of confidence.

As explained our CFD workflow must be parametrized according to the user inputs. To test it in a controlled way we used the solution strategy found to calibrate the CFD for the mono-hedral hull to study the warped hull scaling our workflow using warped hull characteristics.

The percentage differences obtained using the automatic workflow are in reasonable agreement compared also to the expected differences 
for the mono-hedral hull.
Resistance differences are within 5-7\% \textit{wsa} is between 4-13$\%$ and pitch angle is below 5$\%$

\subsection{Automatic parametrization}
From the technical point of view the parametrization is handled by means of a set of Python scripts that works as follows:
\begin{itemize}
	\item Load the \textit{lincosim.inp} file: this file contains all the end-user inputs in a \textit{json} format.
	\item Read the hull geometry file: this file is read to get the hull bounding box extrema.
	\item Compute parameters values: by solving a set of algebraic equations we are able to compute all the parameters of our workflow with actual values.
	\item Write actual values to the simulation directory: this task is accomplished just writing a set of ASCII files that are included by the OpenFOAM dictionaries.
\end{itemize}

%% file: conclusions.tex
\section{Conclusions and perspectives}
We have presented \textit{LincoSim}, a multi-layered automatic platform to perform top-quality
virtual towing tank analysis simulations and manage them so that the results can be 
conveniently part of a design process workflow. The user interface of the platform 
is entirely web-based and does not require a specific expertise on CFD side nor on HPC 
usage side. Preparation, run, post-processing of simulations are automatically performed
so that the user can safely interact with the results as visualizations as well as 
with database for analyses and comparisons. The platform has been extensively tested and 
the quality of results have been assessed during a thorough validation stage based on 
existing experimental data. 

The \textit{LincoSim} platform is currently under testing by the naval companies involved
in LINCOLN. Given further feedbacks from many realistic simulations
we expect to gain confidence on the quality of the results in different use cases and 
possibly to improve the existing simulation setups. A complete validation will
take advantage of the experiments conducted by the naval companies as scheduled in the
LINCOLN project. We will also consider
the possibility to add simulation setups based on different CFD solvers (even 
commercial ones) and compare the quality and time-to-results indicators.
On the other side, we plan to support -- dynamically created --
virtual machines as computing machines in addition to HPC clusters.
Anyhow, the results obtained so far prove that the \textit{LincoSim} integration of well established
tools -- web-server, database, visualization libraries, CFD meshers, CFD solvers, etc. -- providing
a single access-point to the users -- the web browser -- is a significant step forward
to minimize the effort required to perform top-quality naval simulations. Moreover, the
standardized data, operation and result view are a viable way to adequately managing
single and sets of simulations. This approach can dramatically improve the naval design loop evaluation.

In future evolutions, the \textit{LincoSim} platform can be extended to include further
aspects of naval design by integrating, for example, even cycles of automatic
design optimization. In this context, thanks to the standardized approach on data and the
operations management through web-services, \textit{LincoSim} can also be integrated into
other existing frameworks for managing complex workflows (see e.g., \cite{ludascher2006}).

A second line of development concerns extending \textit{LincoSim}'s approach to others
engineering fields that still require very specific skills for execution
and analysis of the results of the corresponding `` virtual experiments ''. 
In this respect, naval design share important challenges with other engineering fields.

%% file: credits.tex
\begin{acknowledgements}
This work is part of the LINCOLN project which has received funding from the European Research Council (ERC) under the European Union's Horizon 2020 research and innovation programme (grant agreement n° 7279827).
We also acknowledge Professor Ermina Begovic and Professor Carlo Bertorello from \textit{Univeristà Federico II di Napoli, Department of industrial engineering} who provided hull CAD files and supported us to complete the CFD worflow validation procedure.
\end{acknowledgements}